\documentclass[a4paper,fleqn]{cas-dc}

\usepackage[numbers]{natbib}

\def\tsc#1{\csdef{#1}{\textsc{\lowercase{#1}}\xspace}}
\tsc{WGM}
\tsc{QE}
\tsc{EP}
\tsc{PMS}
\tsc{BEC}
\tsc{DE}
\begin{document}
\let\WriteBookmarks\relax
\def\floatpagepagefraction{1}
\def\textpagefraction{.001}
\shorttitle{Monopole and quadrupole modes in Mo isotopes}
\shortauthors{CGKNP}

\title[mode = title]{Isoscalar monopole and quadrupole modes in Mo isotopes: microscopic analysis}
%\tnotemark[1,2]

%\tnotetext[1]{This document is the results of the research
%              project funded by the National Science Foundation.}

%\tnotetext[2]{The second title footnote which is a longer text matter
%              to fill through the whole text width and overflow into
%              another line in the footnotes area of the first page.}

\author[1,2]{Gianluca Col\`o}[orcid=0000-0003-0819-1633]
%\cormark[1]
%\fnmark[1]
%\ead{colo@mi.infn.it}
%\ead[url]{www.mi.infn.it/~colo}
\address[1]{Dipartimento di Fisica ``Aldo Pontremoli'', Universit\`a degli
            Studi di Milano, via Celoria 16, I-20133 Milano, Italy}
\address[2]{INFN, Sezione di Milano, Via Celoria 16, I-20133 Milano, Italy}

\author[3,4]{Danilo Gambacurta}
%\ead{gambacurta@lns.infn.it}
\address[3]{Extreme Light Infrastructure - Nuclear Physics (ELI-NP),
Horia Hulubei National Institute for Physics and Nuclear Engineering,
30 Reactorului Street, RO-077125 Magurele, Jud. Ilfov, Romania}
\address[4]{LNS-INFN, I-95123, Catania, Italy}

\author[5]{Wolfgang Kleinig}
%\ead{kleinig@theor.jinr.ru}
\address[5]{Laboratory of Theoretical Physics, Joint Institute for Nuclear Research, Dubna, Moscow region, 141980, Russia}

\author[6]{Jan Kvasil}
%\ead{kvasil@ipnp.troja.mff.cuni.cz}
\address[6]{Institute of Particle and Nuclear Physics, Charles University, CZ-18000,
            Praha 8, Czech Republic}

\author[5,7,8]{Valentin O. Nesterenko}
%\ead{nester@theor.jinr.ru}
\address[7]{State University ``Dubna'', Dubna, Moscow Region, 141980, Russia}
\address[8]{Moscow Institute of Physics and Technology, Dolgoprudny, Moscow region, 141701, Russia}

\author[9]{Alessandro Pastore}
%\ead{alessandro.pastore@york.ac.uk}
\address[9]{Department of Physics, University of York, Heslington, York, YO10 5DD, UK}

\begin{abstract}
The recent RCNP $(\alpha, \alpha')$ data on the Isoscalar Giant Monopole Resonance (ISGMR) and Isoscalar
Giant Quadrupole Resonance (ISGQR) in $^{92,94,96,98,100}$Mo are analyzed within a fully self-consistent
Quasiparticle Random Phase Approximation (QRPA) approach with Skyrme interactions, in which pairing
correlations and possible axial deformations are taken into account. The Skyrme sets SkM*, SLy6,
SVbas and SkP$^{\delta}$, that explore a diversity of nuclear matter properties, are used. We
discuss the connection between the line shape of the monopole strength ISGMR and the deformation-induced
coupling between the ISGMR and the $K=0$ branch of the ISGQR. The ISGMR centroid energy is best described
by the force SkP$^{\delta}$, having a low incompressibility $K_{\infty}$ = 202 MeV. The ISGQR data are
better reproduced by SVbas, that has large isoscalar effective mass $m^*/m$ = 0.9.
The need of describing simultaneously the ISGMR and ISGQR data is stressed, with the requirement of
suitable values of $K_\infty$ and $m^*/m$. Possible
extensions of the QRPA to deal with soft systems are also envisaged.
\end{abstract}

%\begin{graphicalabstract}
%\includegraphics{figs/grabs.pdf}
%\end{graphicalabstract}

%\begin{highlights}
%\item Research highlights item 1
%\item Research highlights item 2
%\item Research highlights item 3
%\end{highlights}


\maketitle

\section{Introduction}

There is still a high theoretical and experimental interest
in the determination of the parameters of the Equation of State (EoS) of nuclear
matter (NM) \cite{ROCAMAZA201896}. Among these
parameters, the nuclear incompressibility $K_\infty$ and the isoscalar effective
mass $m^*/m$ constitute crucial benchmarks for testing new models
and provide an indispensable guideline for applications of nuclear theory to
heavy-ion collisions \cite{GIULIANI2014116}, astrophysical processes
\cite{RevModPhys.74.1015,Burrows2018}, and other areas.

The incompressibility of symmetric NM (SNM) is defined as
\begin{equation}\label{eq:incompr}
K_\infty = 9\rho_0^2 \frac{d^2}{d\rho^2} \left( \frac{E}{A} \right)_{\rho=\rho_0},
\end{equation}
where $E/A$ is the energy per particle and $\rho_0$ is the saturation density
at which the EoS displays a minimum ($\rho_0=0.16$ fm$^{-3}$). Being related
to the second derivative of the EoS around this minimum, $K_\infty$ measures the stiffness of
SNM with respect to the compression. It can be linked to compressional
modes of finite nuclei, in particular to the isoscalar giant monopole resonance
(ISGMR), a breathing mode characterised by a strong transition amplitude from the ground-state
\cite{Blaizot1980,GC.2018}.
Indeed the ISGMR is the main, although indirect, source of information on $K_\infty$.

The discussion on how to extract $K_\infty$ from the ISGMR dates back to the
years 1980s \cite{Blaizot1980} (for the present state-of-the-art, cf.
the recent review \cite{GC.2018}). Nowadays, there is a general consensus
that the relationship between the ISGMR energy and $K_\infty$ may be only
obtained within the self-consistent Energy Density Functional (EDF) theory
\cite{Schunck,APX}. As shown by various EDF calculations,
in nuclei like $^{90}$Zr or $^{208}$Pb the computed ISGMR energy correlates well with the
value of $K_\infty$ of the given functional. This allows, at least in principle, to
consider as correct the $K_\infty$ value associated with the EDF which reproduces
the ISGMR experimental data.

However, the exploration of ISGMRs and their relation to $K_\infty$ still has
some unresolved problems. While $K_\infty$ extracted from $^{208}$Pb
is around 240 $\pm$ 20 MeV, its value in  Sn and other open-shell nuclei is
lower by $\approx$10\% i.e. open-shell nuclei show up some softness
against the compression, see e.g. \cite{Avogadro,fluffy3,PhysRevC.82.024322}.
This can be partly explained by pairing correlations in open-shell nuclei, which somewhat
shift the ISGMR towards lower energies due to the attractive character of the pairing force
\cite{fluffy3}. However, the quantitative results depend to some extent on the choice
of the parameters of that force \cite{PhysRevC.82.024322}. So this problem, often referred
to as "why open-shell nuclei are soft", still remains open.

ISGMR in deformed nuclei deserves a special attention. When a nucleus displays
an axial deformation in its intrinsic frame, the total angular momentum $J$ is no
longer a good quantum number and the nuclear states are characterised
by the projection $K$ of the angular momentum on the intrinsic symmetry axis.
In general such states are superpositions of contributions
from different $J$ of the same parity. In particular, $K=0$ monopole states are
coupled to $K=0$ quadrupole (hexadecapole, etc.) states, which can lead
to an appreciable mixing of the ISGMR and $K=0$ ISGQR. Then,
the monopole strength is redistributed and, in addition to the main ISGMR,
there appears a minor low-energy monopole branch at the same energy as the $K=0$ ISGQR
branch. This is the deformation-induced splitting of the ISGMR.

Following early macroscopic models \cite{ABGRALL1980431,JANG1983303}, and a recent
self-consistent Skyrme Quasiparticle Random Phase Approximation (QRPA)
\cite{PhysRevC.94.064302} analysis, the ISGMR/ISGQR mixing is not complete and
still preserves the main character of each resonance. In addition, the deformation can
shift the energy of the main ISGMR peak (upward in prolate nuclei and downward in oblate nuclei)
\cite{ABGRALL1980431,PhysRevC.94.064302}. This deformation effect, being
distinctive in well-deformed nuclei, becomes more subtle in nuclei with moderate
deformation (i.e., for $\beta <$ 0.2, where $\beta$ is the quadrupole deformation
parameter) \cite{PhysRevC.94.064302}. A further microscopic study of this point
is desirable, e.g. like that for Nd and Sm isotopes in Ref. \cite{YoshidaNakPRC13}.
Such study obviously calls for EDFs that reproduce equally well the monopole and quadrupole strengths.
While the ISGMR peak energy correlates with $K_\infty$, the ISGQR energy is
sensitive to the nucleon isoscalar effective mass $m^*/m$ \cite{Blaizot1980,Nest08}.
The values $K_\infty$ and $m^*/m$ are coupled in SNM, at least for Skyrme EDFs
in which $m^*/m$ is not 1 \cite{ChabanatNPA97}. 
This is an additional argument to consider the ISGMR and ISGQR simultaneously.

Inelastic $\alpha$-scattering data for $^{92,94,96,98,100}$Mo at incident laboratory energy
E$_\alpha$ = 386 MeV have been recently collected at RCNP with the aim to clarify
the issue of how compressible are open-shell (deformed) nuclei \cite{tbp}. Previous
experimental data obtained at TAMU for the Mo isotopes have been published in
\cite{PhysRevC.88.021301,PhysRevC.92.014318}. Soft Mo isotopes seem to be ideal candidates to
discuss the open questions introduced above.

In this study, we analyze the RCNP data within fully self-consistent
Skyrme QRPA models. In Ref. \cite{tbp}, these data have been presented
in conjunction with simple RPA calculations that do not account for the existence of
pairing and possible deformation in the Mo isotopes (with the associated monopole-quadrupole
coupling). Consequently, here we
adress the questions: (i)
how do pairing and deformation shape the ISGMR and ISGQR strengths in soft medium-mass
deformed nuclei? and (ii) how does this impact on our understanding of nuclear incompressibility and
isoscalar effective mass?

\section{Formalism}

The calculations are performed using two different QRPA methods based
on Skyrme EDFs. Both methods are fully self-consistent, i.e. their mean field
and residual interaction are derived from the initial functionals without approximations.
Axial symmetry is assumed. The agreement
between results of these two methods strengthens our conclusions.

\begin{table}[width=.9\linewidth,cols=4,pos=h]
\caption{Incompressibility $K_\infty$ and isoscalar effective mass $m^*/m$
for the Skyrme forces SVbas, SkM$^*$, SLy6, and SkP$^\delta$.}\label{table1}
\begin{tabular*}{\tblwidth}{@{} LLLLL@{} }
\toprule
                & SVbas & SLy6 & SkM$^*$ & SkP$^\delta$ \\
\midrule
$K_\infty$ [MeV]& 234 & 230   & 217  & 202 \\
$m^*/m$       &  0.9& 0.69  & 0.79 & 1  \\
\bottomrule
\end{tabular*}
\end{table}

We employ a representative set of Skyrme forces (SkM$^*$
\cite{Bartel1982}, SLy6 \cite{Cha.98}, SVbas \cite{KlupfelPRC} and SkP$^\delta$
\cite{Dobaczewski_1995}) that span different values of $K_{\infty}$ and $m^*/m$ (see
Table \ref{table1}). The force SkP$^\delta$ may produce instabilities
\cite{Hellemans:2013} but it is employed to include the case of low $K_\infty$
and large $m^*/m$. Giant resonances are less sensitive to
such instabilities, although these may show up when the basis size is increased
\cite{PhysRevC.92.024305}.

The first method, called here as QRPA-I, is introduced in Refs.
\cite{Repko2017,repko2015skyrme}. In this method, the nuclear mean
field and pairing field are computed with the code SKYAX \cite{skyax}
using a two-dimensional mesh in cylindrical coordinates. In our particular
case, the box in which the nuclei are confined extends up to three times the nuclear radii, and
the mesh size is 0.4 fm.
Pairing correlations are included at the level of the iterative HF-BCS (Hartree-Fock plus
Bardeen-Cooper-Schrieffer) method \cite{Repko2017}. We use volume pairing
for SkM$^*$, SLy6, and SkP$^\delta$, and density-dependent pairing for SVbas.
The proton and neutron pairing strengths are fitted to reproduce empirical
pairing gaps obtained from the five-point formula along selected isotopic
and isotonic chains \cite{KlupfelPRC}. To cope with the divergent character of zero-range pairing
forces, energy-dependent cut-off factors are used.
QRPA is implemented in the matrix form. The two-quasiparticle (2qp)
configuration space extends up to 80 MeV, which allows to exhaust the isoscalar E0 and E2
energy-weighted sum rules. The pairing-induced spurious admixtures are
extracted following the prescription of \cite{Repko2019SEARPA}.

The second method, called as QRPA-II, follows closely Ref. \cite{Losa}.
The Hartree-Fock-Bogoliubov (HFB) equations are solved in an axially
deformed harmonic oscillator (HO) basis by using the code HFTHO \cite{Stoitsov}.
A canonical basis with 14 major HO shells is used.
As additional cut-off, canonical states with energies larger than 200 MeV
or pairs with occupation factors $v^2 < v^2_{\rm crit}=10^{-2}$ are discarded.
For each Skyrme force, the pairing strength is fixed so that the canonical
neutron pairing gap $\Delta_n$ = 1.4 MeV in $^{120}$Sn is reproduced.
QRPA eigenvalues and eigenvectors are found by using diagonalization techniques for
sparse matrices.

The implementation of various prescriptions in the cases I and II allows us
to conclude that: (i) the inclusion of pairing is mandatory, and (ii)
the details of the procedure (HFB vs. HF-BCS) or of the pairing force are not
crucial, as far as the resulting pairing gaps are close to the empirical ones. In fact,
all our models lead to similar results in the Mo isotopes: proton gaps are slightly larger
than neutron gaps but all are around 1 MeV. Fine details will be
reported in a forthcoming publication.

\begin{figure*}
\centering
\includegraphics[scale=.30]{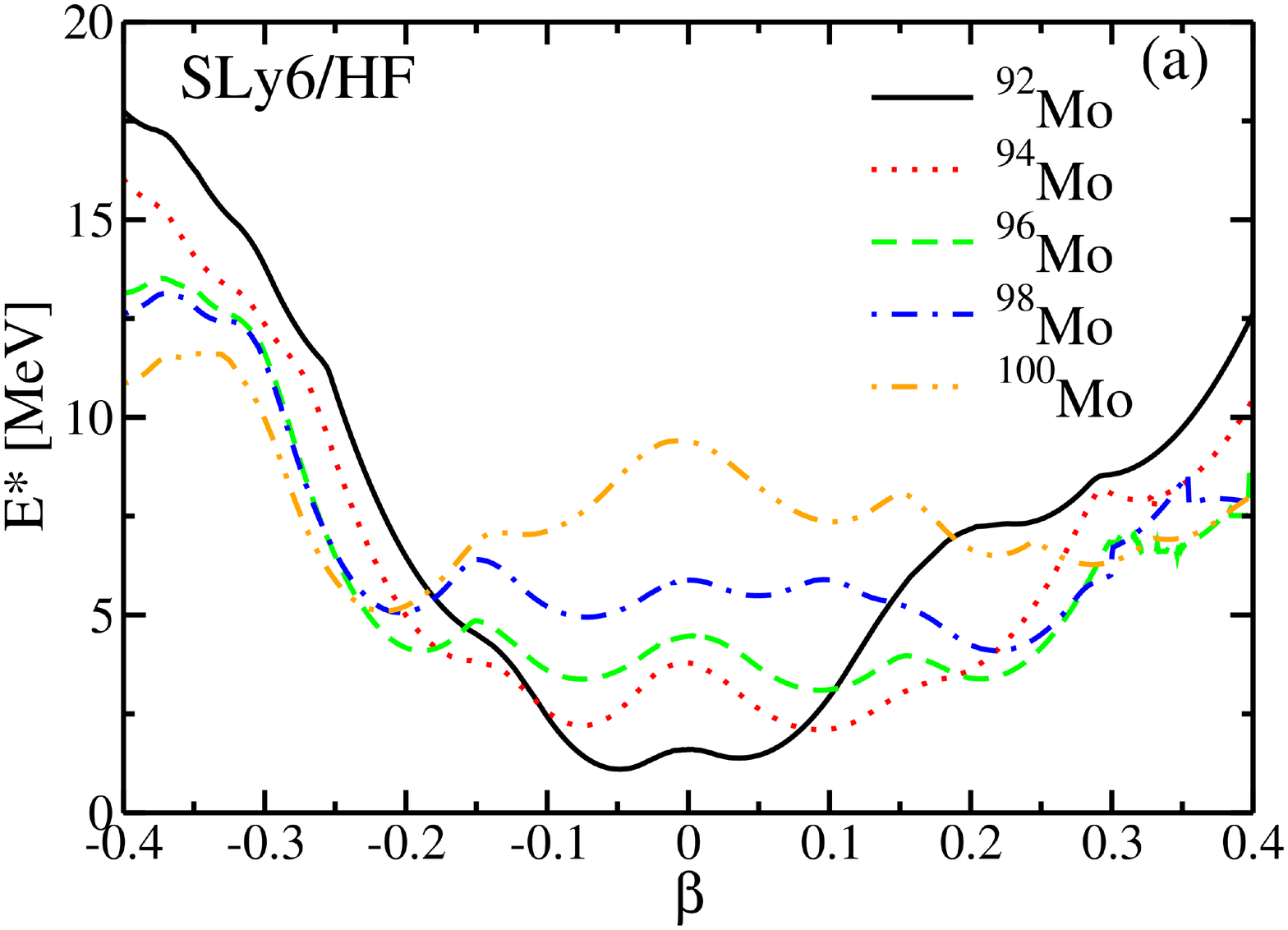}
\includegraphics[scale=.30]{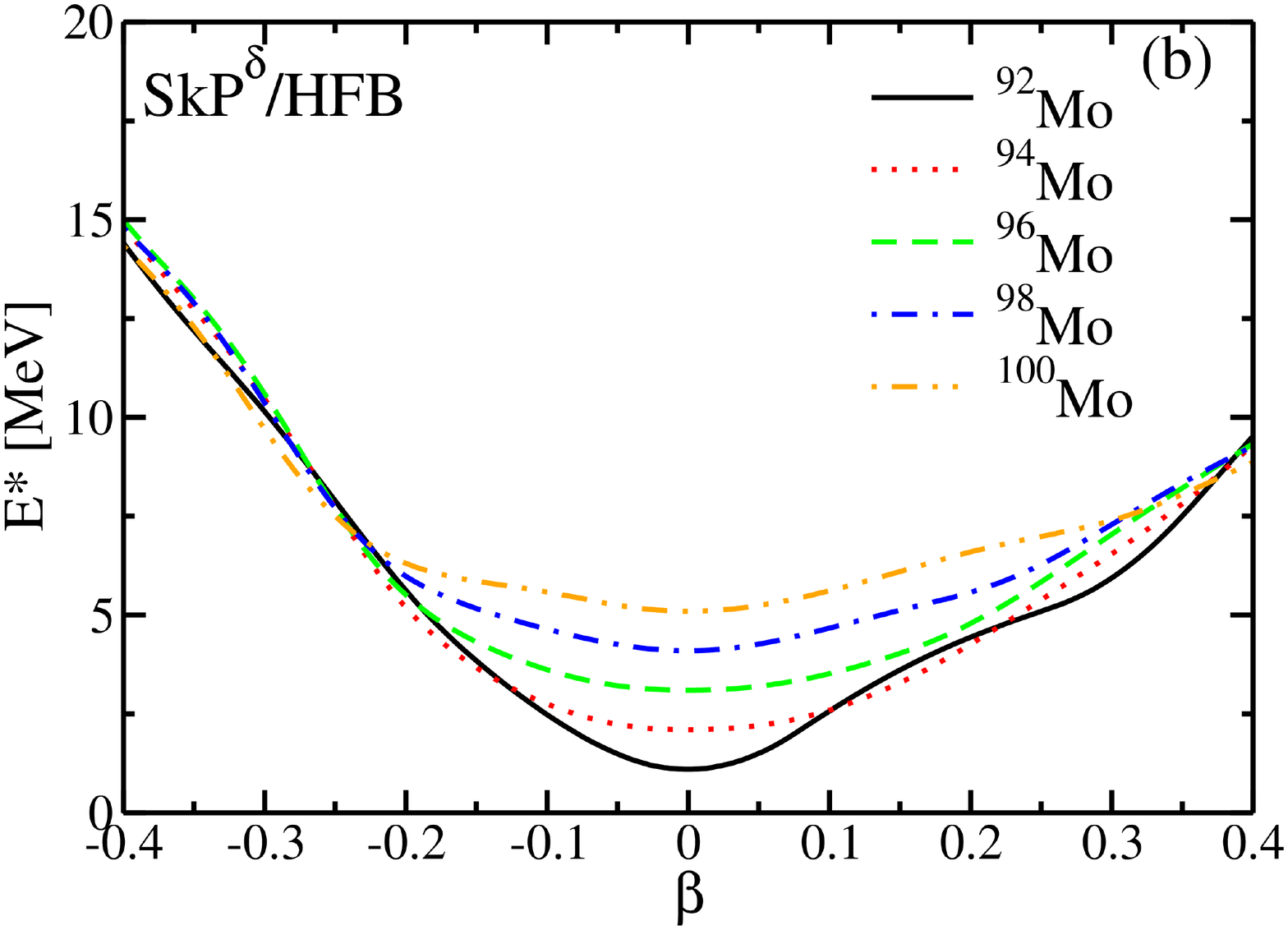}
\includegraphics[scale=.30]{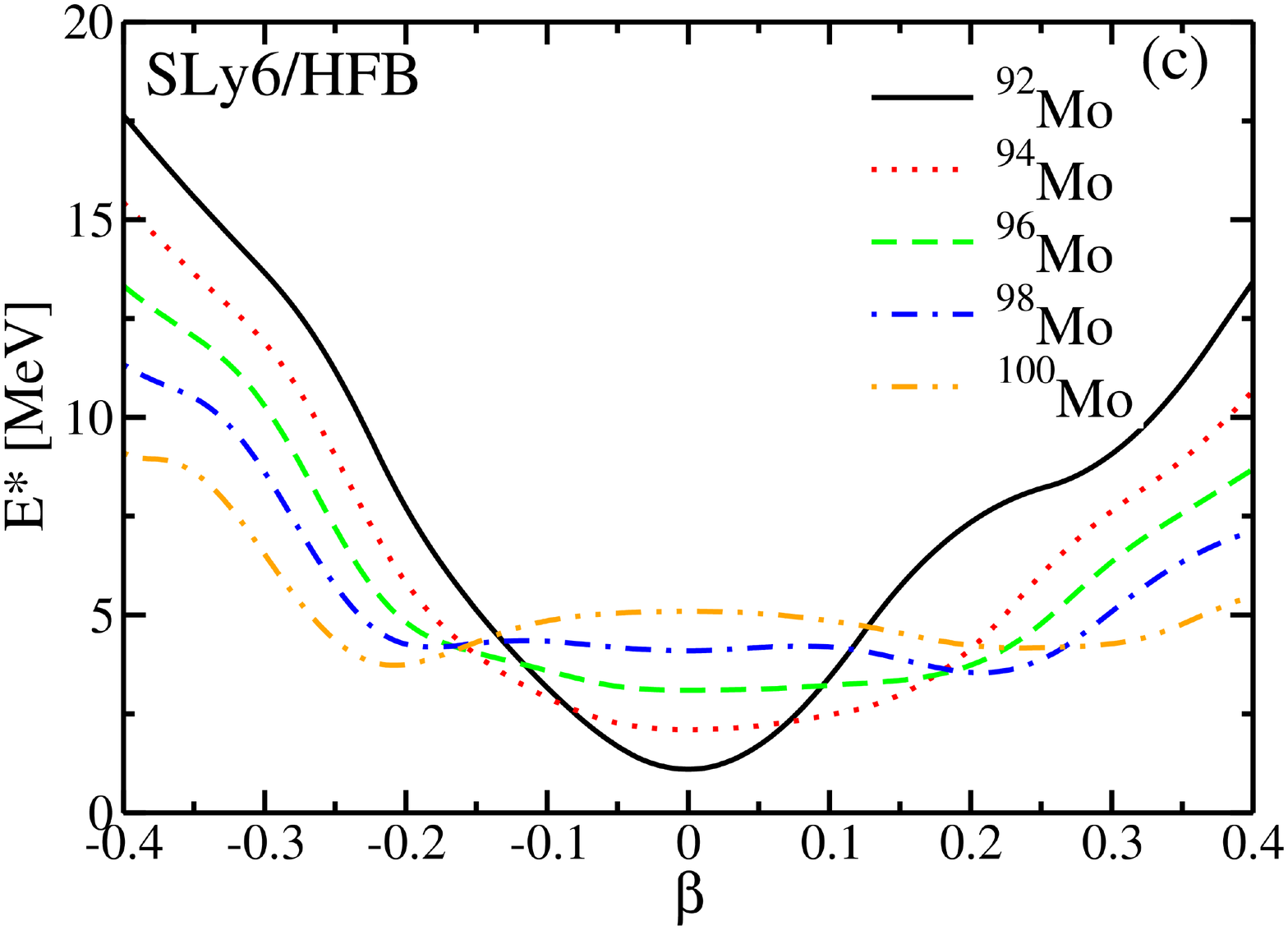}
\includegraphics[scale=.30]{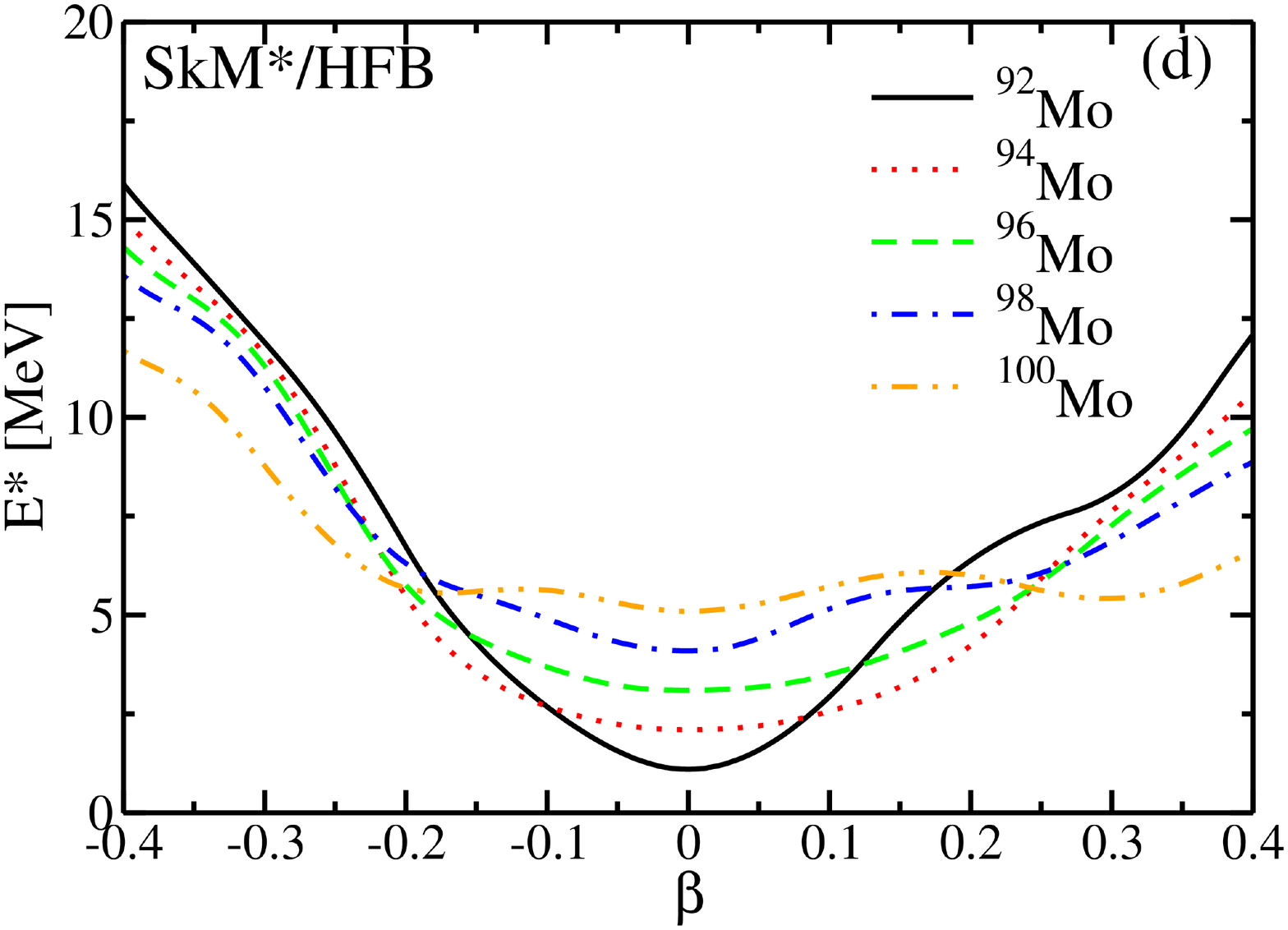}
\caption{Potential energy curves obtained by constrained calculations with different Skyrme forces.
In the left panels, we report results obtained with SLy6 at the level of HF (upper panel) and
HFB (lower panel). In the upper right (lower right) panels, HFB results associated with the
SkP$^\delta$ (SkM$^*$) functional are displayed.}
\label{FIG:1}
\end{figure*}

\section{Results for the Mo isotopes}
\subsection{Ground-state deformation and softness}

In this subsection we discuss the calculated potential energy curves (PECs)
for various Skyrme EDFs. Though some Mo isotopes may exhibit triaxiality
\cite{Donau07}, we restrict ourselves, for the sake of simplicity, to the case
of axial deformation characterised by the deformation parameter $\beta$. Each
PEC point is obtained by minimization of the total ground-state
energy under the constraint of a fixed $\beta$ (and corresponding quadrupole moment).

In Fig. \ref{FIG:1} we display the PECs obtained by our method II for
SkM$^*$, SLy6, and SkP$^\delta$. The comparison of SLy6 results obtained by HF
(upper left panel) and HFB (lower left panel) shows that
pairing has the important effect of smoothing the PECs. This can
be probably attributed to the fact that pairing correlations smear out the
occupancies and so make the total energies less
sensitive to details of the deformed single-particle levels.

Fig. \ref{FIG:1} shows that PECs associated with different EDFs, albeit not identical,
lead to the same qualitative outcome. Pairing makes the spherical minima
more favoured but, overall, Mo isotopes look soft against quadrupole deformation.
The softness increases with the neutron number. Even in $^{92}$Mo the potential is
not steep although the spherical minimum is well defined. In the heavier isotopes,
the potential is increasingly shallow and the spherical minimum is not well defined at all.
The extreme case of a flat potential is reached for $^{98}$Mo and $^{100}$Mo.
In the case of SLy6, $^{100}$Mo displays even a kind of convex potential.
We do not plot PECs for SVbas, but we have checked that they are consistent with all
the above conclusions.

It is not easy to obtain a straight experimental confirmation of ground-state (g.s.)
deformation in soft nuclei. The isotopes $^{94,96,100}$Mo exhibit
g.s. rotational bands with $J^{\pi} = 0^+, 2^+, 4^+, 6^+, 8^+$, \cite{nndc.bnl}
and so they should have at least a modest axial quadrupole deformation. The values of
the deformation parameter, extracted from E2 transitions in the g.s. band, are
$\tilde{\beta}$ = 0.109, 0.151, 0.172, 0.168,
0.162 for A=92, 94, 96, 98, 100, respectively \cite{nndc.bnl}. However, in soft
nuclei, the deformation parameters obtained in such a way can be overestimated.
This is confirmed by the fact that the energies
of the first 2$^+$ states in $^{94,96,98}$Mo, calculated with $\tilde{\beta}$,
are lower than the experimental values. In summary, the values $\tilde{\beta}$
should be rather considered as upper limits of the
true deformation parameters.

In keeping with all these facts, we performed QRPA calculations for
the Mo isotopes both on top the self-consistent minimum (which is spherical
in most of the cases), and with the constraint $\beta=\tilde{\beta}$. This
provides us a proxy for the softness and theoretical uncertainties in our
quest on how the strength distributions are sensitive to a modest axial deformation.

\subsection{QRPA strength distributions}

In this subsection we discuss our main results, namely, QRPA strength distributions. The
isoscalar monopole (L=0) and quadrupole (L=2) strengths are defined in the usual way as
\begin{equation}\label{eq:strength}
S_L(E) = \sum_{K=0}^L (2-\delta_{K,0}) \sum_{\nu \in K}
\vert \langle \nu \vert {\hat O}_{LK} \vert 0 \rangle \vert^2
\xi_{\Delta}(E-E_\nu),
\end{equation}
where
$\nu$ labels the complete set of QRPA eigenvalues $|\nu\rangle$
with the energies $E_{\nu}$,
$|0\rangle$ is QRPA ground-state,
and the monopole and quadrupole isoscalar transition operators
read $\hat O_{00} = \sum_i^A r^2_i Y_{00} (\hat r_i)$
and $\hat O_{2K} = \sum_i^A r^2_i Y_{2K} (\hat r_i)$.
For a more convenient comparison with experimental data, the strength is
smoothed by the Lorentz weight $\xi_{\Delta}(E-E_{\nu}) = \Delta /(2\pi
[(E-E_{\nu})^2 - \Delta^2/4])$, with the averaging parameter $\Delta$ = 2.5 MeV.

\begin{figure}
\centering
\includegraphics[height=7cm,width=8cm]{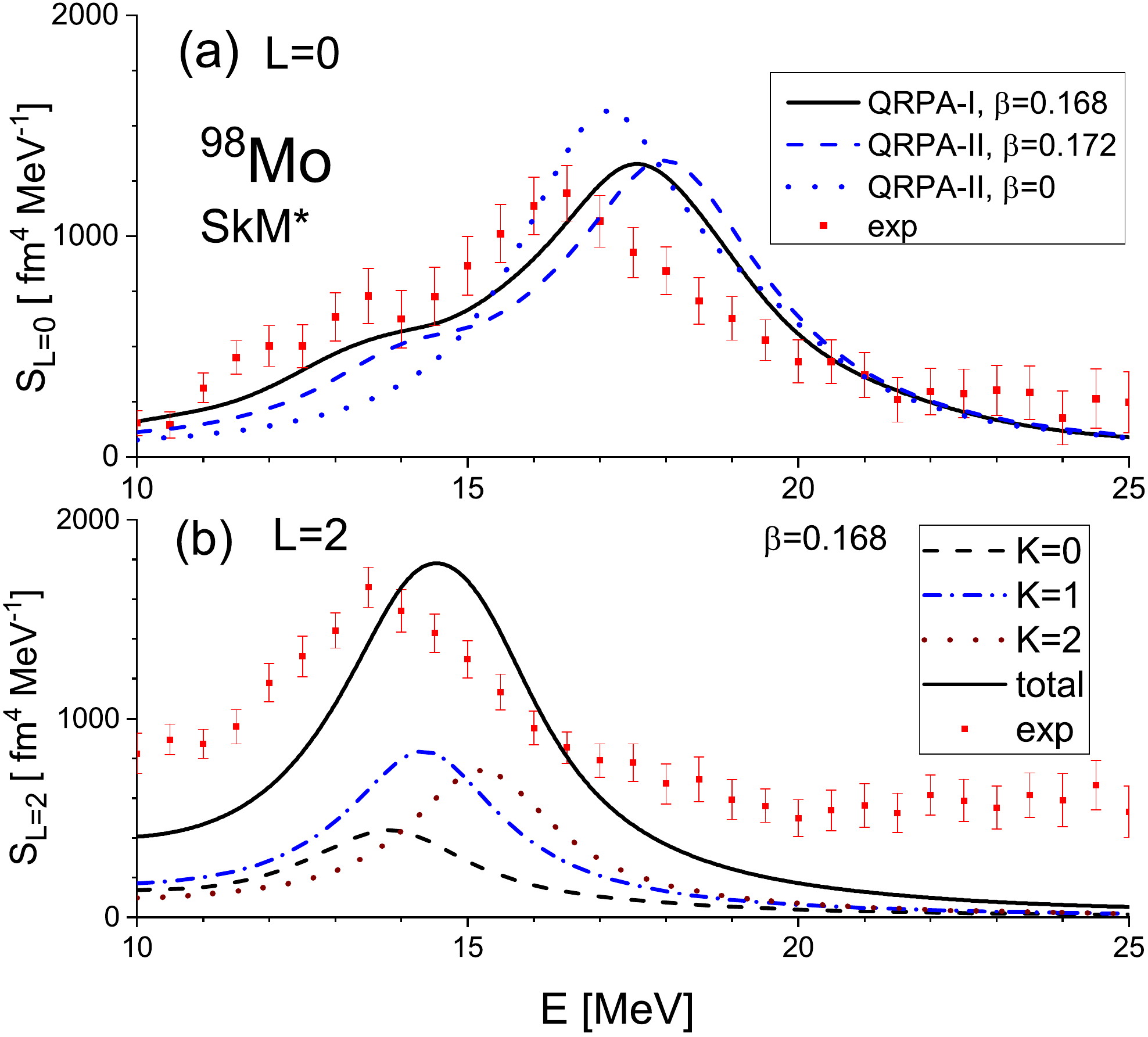}
\caption{a) Monopole (L=0) strengths in $^{98}$Mo, calculated within QRPA-I and QRPA-II.
b) Quadrupole (L=2)
strengths in $^{98}$Mo, calculated within QRPA-I.
In both panels, the experimental data from \cite{tbp} are used.}
\label{FIG:2}
\end{figure}

Before a general discussion of ISGMR and ISGQR along the Mo isotope chain,
we would like to highlight some important points. They are illustrated in Fig. \ref{FIG:2}
for the SkM$^*$ strengths in $^{98}$Mo.

1) In panel (a), the monopole strengths obtained by QRPA-I and QRPA-II at
close deformations $\beta$ = 0.168 and 0.172 are compared.
Though the implementations are slightly different,
the QRPA-I and QRPA-II curves are pretty consistent with each other.
The somewhat higher ISGMR peak energy in QRPA-II can be partly explained by
the larger deformation.

2) By comparing in panel (a) two QRPA-II results with different g.s. deformations
($\beta$ = 0 and 0.172), we clearly see that even a modest deformation can significantly shift upward
the ISGMR peak energy, by $\approx$ 0.8 MeV in this case. Moreover, the ISGMR has a single peak
in the spherical case ($\beta$ = 0), and it acquires a noticeable low-energy shoulder at
deformation $\beta$ = 0.172.

3) In panel (b), the deformation splitting of the ISGQR into branches with $K=0, 1, 2$
is shown. It is easy to see that the ISGMR shoulder of QRPA-I in panel (a) and the $K=0$
branch of the ISGQR in panel (b) lie in the same energy interval. This confirms that the
shoulder arises due to the deformation-induced coupling of the monopole and quadrupole modes.

We should also note that, from Fig. \ref{FIG:2}, SkM* significantly overestimates both ISGMR and ISGQR
experimental peak energies. This indicates that the values $K_{\infty}$ = 217 MeV
and $m^*/m$ = 0.79 are not optimal. For a better reproduction of the experimental data,
smaller (larger) values of $K_{\infty}$ ($m^*/m$) are desirable.

\begin{figure*}
\centering
\includegraphics[width=15cm]{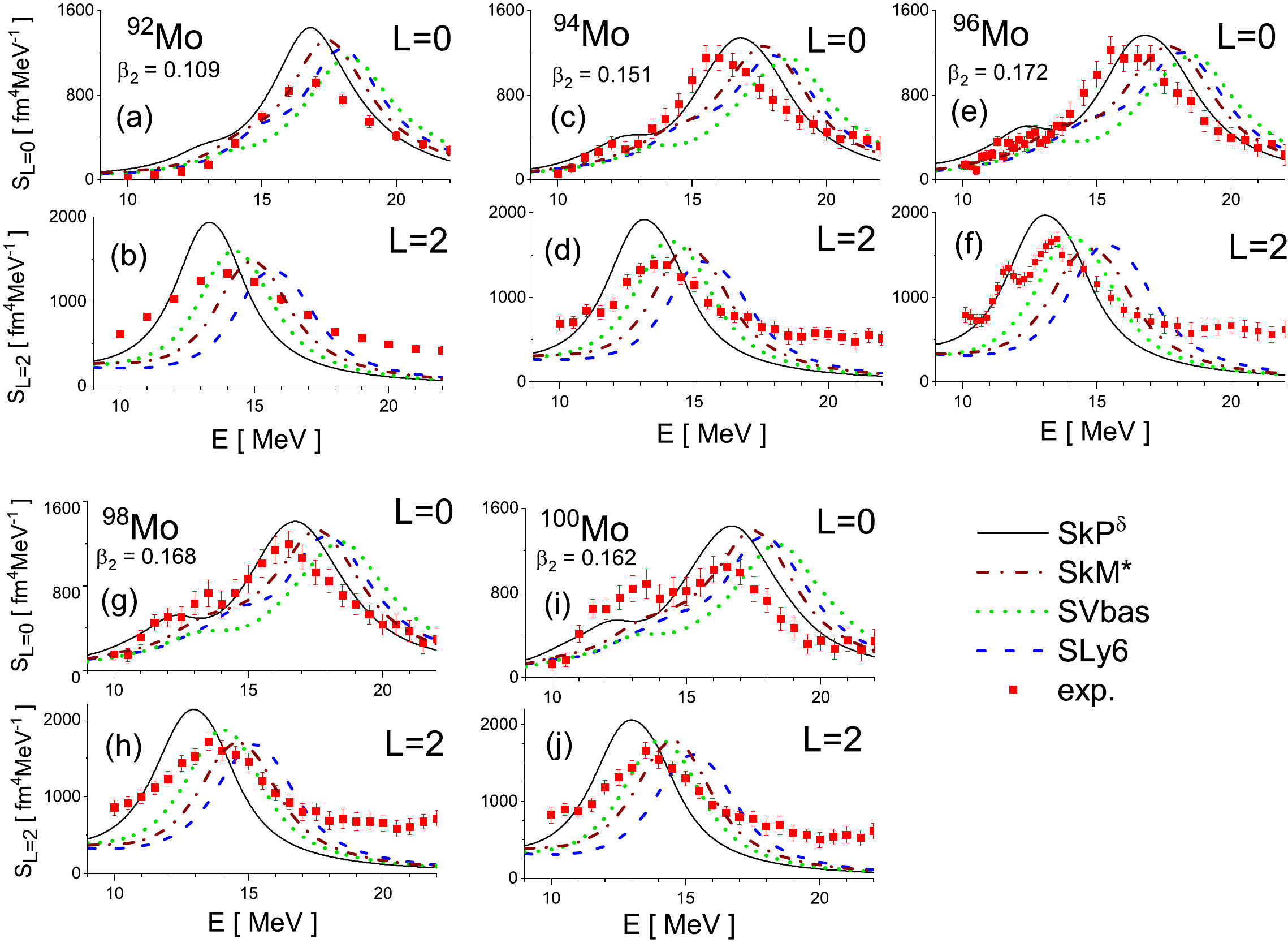}
\caption{Isoscalar monopole (L=0) and quadrupole (L=2) strengths in $^{92,94,96,98,100}$Mo, calculated
in the framework of QRPA-I with the Skyrme forces SkM$^*$, SLy6, SVbas and SkP$^{\delta}$. The strengths
are compared with the experimental data \cite{tbp}.}
\label{FIG:3}
\end{figure*}

In Fig. \ref{FIG:3}, we provide a general view on monopole and quadrupole strength
along the isotopic chain. QRPA-I results obtained with different Skyrme forces
at the deformations $\beta=\tilde{\beta}$ (see above) are compared with the
experimental data \cite{tbp}. The data show that, in addition to the main ISGMR peak,
a low-energy shoulder appears in $^{94,96,98}$Mo. This structure resembles more a double hump
in $^{100}$Mo. The larger the deformation of the isotope, the more pronounced the
shoulder. As discussed in connection with Fig. \ref{FIG:2}, the shoulder is caused
by the deformation-induced monopole-quadrupole coupling. So one may conclude that
this coupling can manifest itself even at a modest deformations, $\beta$ = 0.15-0.18.

Fig. \ref{FIG:3} shows that SVbas ($K_\infty$ = 234 MeV), SLy6 ($K_\infty$ = 230 MeV), and
SkM*($K_\infty$ = 217 MeV) overestimate the experimental ISGMR peak energies by 1.4-2.8 MeV,
1.0-2.2 MeV and 0.6-1.8 MeV, respectively.
For all these three forces, the overestimate is minimal in $^{92}$Mo and maximal in $^{94}$Mo.
SkP$^\delta$ ($K_\infty$ = 202 MeV) reproduces the ISGMR data much better: the overestimate
of the peak energies are 0.0 MeV ($^{92}$Mo), 0.8 MeV ($^{94}$Mo), 0.5 MeV ($^{96}$Mo),
0.2 MeV ($^{98}$Mo), and 0.1 MeV ($^{100}$Mo). Thus
Mo isotopes call for a quite low value of $K_\infty$. However, the additional strong
effect of the monopole-quadrupole coupling prevents us from a quantitative conclusion.

Moving to the quadrupole strengths in Fig. \ref{FIG:3}, we first of all notice
that the experimental data of \cite{tbp} do not show fine structures in all isotopes,
aside from $^{96}$Mo (that has the largest $\tilde\beta$). As discussed in Ref.
\cite{KUREBA2018269}, deformation mainly
produces a broadening of the ISGQR. This is confirmed by Fig. \ref{FIG:2}, where the
splitting of the ISGQR into $K$-components is shown to be too small to distinguish
the separate $K$-branches in the total strength distribution.

As already mentioned, the ISGQR peak energy correlates with the effective mass
$m^*/m$ \cite{Nest08} or, more precisely, with $\sqrt{m^*/m}$
\cite{Blaizot1980}. From Fig. \ref{FIG:3}, SLy6 and SkM$^*$ with their low effective
masses (0.79 and 0.69, respectively) significantly overestimate the ISGQR peak energy.
Instead, SkP$^\delta$ with $m^*/m$ = 1 systematically underestimate it. SVbas with
$m^*/m$ = 0.9 gives the best agreement: its ISGQR peak energies
overestimate the experimental data only by 0.1-0.5 MeV. In Ref. \cite{Nest08},
it has been noted that a larger value of the effective mass, $m^*/m >$ 0.9,
would spoil the description of the isovector GDR. So
a value of $m^*/m \approx$ 0.9 seems to be optimal.

It is interesting at this stage to note that, within the standard Skyrme framework,
there is an implicit relationship between ISGMR and ISGQR.
In fact, in the Skyrme EoS for SNM,
$K_\infty$ and $m^*/m$ are expressed as \cite{ChabanatNPA97}
\begin{equation}
K_\infty = B + C\sigma + D(1-\frac{3}{2}\sigma)\Theta , \; \;
\frac{m^*}{m}=[1+\frac{m\rho_0}{8\hbar^2} \Theta]^{-1},
\end{equation}
where $B$, $C$, and $D$ are simple functions of the saturation density
$\rho_0$ and saturation energy of SNM,
$\sigma$ is the power in the density-dependent $t_3$-term of the Skyrme force, and $\Theta$ is a
simple combination of Skyrme parameters related to momentum-dependence.
As shown in
Ref. \cite{ChabanatNPA97}, the incompressibility and effective mass can indeed correlate with each other. 
Then, the simultaneous description of the ISGMR and ISGQR is not only desirable but also
provides tight constraints on the Skyrme parameters. However, a more general local functional,
or a covariant EDF, may be not display this correlation between incompressibility and effective mass.

\section{Conclusions and perspectives}

New data on monopole and quadrupole strength distributions in
$^{92,94,96,98,100}$Mo have been obtained in a recent $\alpha$-inelastic scattering
experiment at RCNP, Osaka \cite{tbp}. The main purpose of this
paper is the description of these data within state-of-the-art Quasiparticle
Random Phase Approximation (QRPA) methods with Skyrme interactions. To this aim, a relevant set
of Skyrme forces (SkM$^*$ \cite{Bartel1982}, SLy6 \cite{Cha.98}, SVbas
\cite{KlupfelPRC} and SkP$^\delta$ \cite{Dobaczewski_1995}), characterised by
different values of incompressibility $K_\infty$ and isoscalar effective mass $m^*/m$,
is selected.

We show that, in the ground state (g.s.), the inclusion of pairing and the breaking of spherical
symmetry play important roles. With pairing, different EDFs
give similar predictions. In particular, we have shown that: (i) Mo isotopes are generally
soft with respect to quadrupole deformation,
and (ii) this softness increases with the mass number. In the most neutron-rich Mo isotopes,
the potential energy curve (PEC) as a function of the deformation parameter $\beta$ looks very
shallow. The pairing has the important effect of smoothing out the PECs.

Our QRPA calculations based on a slightly deformed g.s. reproduce the shape of the
monopole strength distributions. In $^{94,96,98,100}$Mo, the monopole strength displays the
main peak and a lower-energy shoulder.
Following our analysis, this shoulder is produced by the deformation-induced
coupling of ISGMR and ISGQR. So, even in nuclei with a modest deformation
($\beta$ = 0.15-0.18 in our calculations) this coupling can have a visible effect.

It is also shown that Skyrme forces with $K_\infty$ between
217 and 234 MeV significantly overestimate the ISGMR peak energies, while
SkP$^\delta$ with $K_\infty$ = 202 MeV gives acceptable results that lie
much closer to the data.
For the successful description of the ISGQR in Mo isotopes, we need Skyrme forces with
$m^*/m\approx$ 0.9 like SVbas and, to a lesser extent, SkP$^\delta$.
In general, it is desirable that ISGMR and ISGQR are described
consistently, but this is even more true in deformed nuclei where the two resonances are
coupled.

The conclusion from our calculations that
$K_\infty$ should be smaller than 210 MeV and
$m^*/m\approx$ 0.9 should be taken with reasonable care. One reason is the
monopole-quadrupole coupling that we have emphasised along our paper. In addition,
QRPA is not a reliable theory for soft nuclei.
To deal with the first issue, projection methods like the one in \cite{erler2014collective}
should be considered.
For soft nuclei, theories that account properly for shape coexistence like multi-reference
DFT (see \cite{MR-DFT} and references therein), are an option. However, so far, they have been applied to low-lying
spectroscopy but
not to giant resonances in nuclei with a flat PEC. Because of all these related aspects,
Mo isotopes themselves and extraction of EoS parameters therefrom,
deserve in future a more careful analysis.

\section*{Acknowledgements}

VON and JK thank Prof. P-G. Reinhard for useful discussions and
Dr. A. Repko for the QRPA code. The work
was partly supported by Votruba - Blokhintsev (Czech Republic - BLTP JINR) grant
(VON and JK) and grant of Czech Science Agency, Project No. 19-14048S (JK).
Partial support (GC) of the funding from the European Union's Horizon 2020
research and innovation program, under grant agreement No 654002,
is also gratefully acknowledged. The work of AP was supported by STFC Grant 
No. ST/P003885/1. Part of the calculations have been performed using the 
DiRAC Data Analytic system at the University of Cambridge, operated by 
the University of Cambridge High Performance Computing Service on behalf 
of the STFC DiRAC HPC Facility (www.dirac.ac.uk). This equipment was funded by a 
BIS National E-infrastructure capital grant (No. ST/K001590/1), STFC 
capital Grants No. ST/H008861/1 and ST/H00887X/1, and STFC.

%\appendix
%\section{My Appendix}
%Appendix sections are coded under \verb+\appendix+.

%\verb+\printcredits+ command is used after appendix sections to list
%author credit taxonomy contribution roles tagged using \verb+\credit+
%in frontmatter.

%\printcredits

%% Loading bibliography style file
\bibliographystyle{model1-num-names}
%\bibliographystyle{cas-model2-names}

% Loading bibliography database
\bibliography{paper_mo_4}

\end{document}